%% file: globs.tex
\RequirePackage{fix-cm}
\documentclass{svjour3}                     
\smartqed  

\makeatletter
\def\cl@chapter{\@elt {theorem}}
\makeatother

\usepackage{natbib}

\usepackage{amsmath,amsfonts}
\usepackage{array}
\usepackage[caption=false,font=normalsize,labelfont=sf,textfont=sf]{subfig}
\usepackage{url}
\usepackage{graphicx}
\usepackage{mathptmx}      

\usepackage{hyperref}
\usepackage{cleveref}

\usepackage[T1]{fontenc}
\usepackage[utf8]{inputenc}

\usepackage{listingsutf8, pythonhighlight}
\usepackage{listings, lstautogobble}
\usepackage{xspace}
\usepackage{booktabs}
\usepackage{xcolor}

\lstdefinelanguage{JavaScript}{
    keywords={typeof, new, true, false, catch, function, return, null, catch, switch, var, if, in, while, do, else, case, break},
    keywordstyle=\color{blue}\bfseries,
    ndkeywords={class, export, boolean, throw, implements, import, this},
    ndkeywordstyle=\color{darkgray}\bfseries,
    identifierstyle=\color{black},
    sensitive=false,
    comment=[l]{//},
    morecomment=[s]{/*}{*/},
    commentstyle=\color{gray}\ttfamily,
    stringstyle=\color{red}\ttfamily,
    morestring=[b]',
    morestring=[b]"
}

\lstdefinelanguage{HCL}{
    keywords={path, capabilities, read, write},
    keywordstyle=\color{blue}\bfseries,
    sensitive=false,
    comment=[l]{\#},
    commentstyle=\color{gray}\ttfamily,
    stringstyle=\color{red}\ttfamily
}

\lstdefinelanguage{Go}{
    keywords={func, const, if, else, return, nil, error, string},
    keywordstyle=\color{blue}\bfseries,
    sensitive=false,
    comment=[l]{//},
    morecomment=[s]{/*}{*/},
    commentstyle=\color{gray}\ttfamily,
    stringstyle=\color{red}\ttfamily
}

\lstdefinelanguage{ebnf}{
    morekeywords={::=},
    sensitive=false,
    morecomment=[l]{//},
    morecomment=[s]{/*}{*/},
    escapeinside={(@}{@)},
}
\usepackage{enumitem}
\setlist[itemize]{noitemsep, topsep=3pt}
\newlist{researchquestions}{enumerate}{1}
\setlist[researchquestions]{label*=\textbf{RQ\arabic*}}

\newcommand{\eg}{\hbox{\emph{e.g.}}\xspace}

\newcommand{\gdos}{\hbox{G-DoS}\xspace}

\newcommand{\cf}[1]{\texttt{#1}}

\journalname{}
\begin{document}

\title{An Empirical Analysis of the Glob Ecosystem}

\author{Phyllis Lim         \and
        Connor Adkins       \and
        Mark Marron
}


\institute{Phyllis Lim \at
              Department of Computer Science, University of Kentucky, KY, USA \\
              \email{phyllis.lim@uky.edu}
            \and
            Connor Adkins \at
              Department of Computer Science, University of Kentucky, KY, USA \\
              \email{chandon.adkins@uky.edu}
            \and
            Mark Marron \at
              Department of Computer Science, University of Kentucky, KY, USA \\
              \email{mark.marron@uky.edu}
}

\date{}

\maketitle

\begin{abstract}
Glob patterns, a domain-specific language for structured string matching, are a foundational yet understudied component of modern software development embedded in everything 
from build scripts and configuration files to web server routes. However, this widespread use rests on a fragile foundation. Despite their ubiquity across tools 
and programming languages, a lack of standardization has led to a fragmented ecosystem rife with inconsistent behaviors, security vulnerabilities, and usability pitfalls. 
This paper presents the first empirical study of the glob ecosystem to quantify these challenges and chart a path toward robust solutions. 

Through an analysis of $1,966$ open source projects, $1,355$ Github issues, $444$ CVE reports, and $361$ StackOverflow posts, we systematically map the feature support and 
real-world usage of globs across six common ecosystems that utilize globs. 
Our findings reveal a stark divide between various globbing implementations and developer adoption. In addition, we document inconsistencies that hinder portability, reliability, 
and create security flaws. Our analysis reveals that security vulnerabilities are not corner cases, but a dominant concern, comprising almost a quarter of all developer discussions. 
We propose a path forward through standardization and introduce a formal specification for globs, \textit{GlobSpec}, designed to resolve semantic ambiguities and 
bridge the expressiveness gap between current implementations and developer requirements.
\keywords{Globs, Pattern Matching, Domain Specific Language, Software Engineering}
\end{abstract}

\section{Introduction}
\label{intro}
\input{intro.tex}

\section{Background}
\label{background}
\input{background.tex}

\section{Research Questions}
\label{sec:researchquestions}
\input{rqs.tex}

\section{Methodology}
\label{sec:methodology}
\input{methodology.tex}

\section{Analysis}
\label{sec:analysis}
\input{analysis.tex}

\section{Discussion}
\label{sec:discussion}
\input{discussion.tex}

\section{Threats to Validity}
\label{sec:threats}
\input{threats.tex}

\section{Related Work}
\label{sec:relatedwork}
\input{related.tex}

\section{Conclusion}
This study provides the first comprehensive analysis of globbing features across programming languages, highlighting inconsistencies, limitations, and areas for improvement. By 
examining real-world usage and cross-platform challenges, we identify the need for standardized implementations and enhanced developer tools to address portability and usability issues.
While globbing is a convenient mechanism for matching file paths, multiple vulnerabilities suggest it can become a surprising attack surface. We highlight Glob-Driven Denial-of-Service 
(\gdos) attacks as an emerging class of vulnerabilities where misuse of glob syntax or poor implementation leads to resource exhaustion, memory leaks, or crashes. Additionally, 
this work proposes a standardization for glob languages that unifies and simplifies many challenges present with existing globbing systems.
These findings lay the groundwork for better practices, opportunities for enhancements to the design, and standardization of globbing semantics, ultimately aiming to provide developers 
with robust, portable, and reliable capabilities. 

\section*{Data-Availability Statement}
The data used in this study, including the glob corpus, forum posts, and CVE issues is available via github (\url{https://anonymous.4open.science/r/glob-artifact-AA95}) as an evolving community data set and as a stable 
reproduction package at Zenodo, (\url{https://doi.org/10.5281/zenodo.18462402}).


\section*{Declarations}
\subsection*{Funding:}
The authors did not receive support from any organization for the submitted work.

\subsection*{Ethical approval:}
Not applicable.

\subsection*{Informed consent:}
Not applicable.

\subsection*{Author Contributions:}
Phyllis Lim lead the data collection and analysis. Connor Adkins led the design and implementation 
work on GlobSpec. Mark Marron contributed to the interpretation of results and manuscript preparation.

\subsection*{Data Availability Statement:}
The data used in this study, including the glob corpus, forum posts, and CVE issues is available via github (\url{https://anonymous.4open.science/r/glob-artifact-AA95}) as an evolving community data set and as a stable 
reproduction package at Zenodo, (\url{https://doi.org/10.5281/zenodo.18462402}).

\subsection*{Conflict of Interest:}
The authors declare that they have no conflict of interest.

\subsection*{Clinical Trial Number:}
Clinical trial number: not applicable.

%
%


\newpage

\bibliographystyle{spbasic}
\bibliography{bibliography}

\end{document}

%% file: intro.tex

Globs are a special case of regular languages that are used extensively to manipulate path-like structured strings. They appear with custom support in configuration formats like 
\texttt{.gitignore}, shell command style expansions. Globs are also supported as specialized domain-specific languages (DSLs) in most general-purpose programming languages as 
either via builtin features, \eg Java in 
\texttt{filepath}, or as popular libraries, such as \texttt{glob} in go. These glob patterns are sources for a variety of odd bugs~\cite{gitignorestarstar} and high-impact classes of security 
vulnerabilities including path traversals, which is a top-10 "Most Dangerous Software Weaknesses"~\cite{mitre2024top25}. In addition, as globs are a dialect of regular expressions, they also have many of the same pitfalls and difficulties that 
are known from empirical research into regular expressions~\cite{regexes-are-hard,regex-usage-and-context,regex-comprehension}. Despite this widespread use and these challenges, 
there is limited empirical study on glob pattern syntax, semantics, and usage. 

Globs support simple wildcard characters like \texttt{*} (multi-character match), \texttt{$?$} (single-character match), and character sets (\texttt{[abc]}) to allow flexible matching. 
Enhanced globbing features, such as recursive directory traversal \texttt{(**)}, and hidden file inclusion \texttt{(.*)}, further extend their utility. Many systems also employ 
additional meta-semantics, \eg \texttt{gitignore} files, where there are special rules for leading/trailing slashes and \texttt{(**)} patterns, as well as order dependent control 
for \texttt{!} which excludes from a previous pattern. Globbing implementations vary significantly across languages and ecosystems. These variations result in inconsistent 
behavior~\cite{globinconsistency}, particularly when scripts are moved between environments, or the same glob is run on platforms with (slightly) different semantics. The lack of 
standardization poses a challenge for developers who want to create robust cross-platform file handling~\cite{regex-generalizbility,regex-evolution}. 

To understand the prevalence and feature sets of globs, this paper undertakes a measurement study on the usage of glob patterns in popular ecosystems, including Bash, Python, Node.js, 
Ruby, PHP, and Go. To provide insight into challenges developers face when working with globs, and areas where features may be missing, this paper contains an analysis of data from 
developer platforms including Github, StackOverflow, and CVE reports. Our findings provide the first empirical study on the practical implications of glob use and design. By shedding 
light on the gaps and inconsistencies, this study provides a foundation for standardizing glob pattern semantics and enhancing developer productivity.\\

\noindent
The contributions of this paper are as follows:
\begin{itemize}
\item We present a new polyglot glob dataset consisting of $62,713$ unique globs extracted from $1,966$  software projects written in six popular programming languages, alongside a curated corpus of 2,160 developer discussions from GitHub issues, CVE reports, and Stack Overflow, providing comprehensive quantitative and qualitative data for glob ecosystem analysis.
\item We empirically validate that glob use is widespread, with a core set of shared semantics \emph{but} also expose critical misalignments between language support and developer needs.
\item We identify a set of common pitfalls, and characterize an emerging form denial-of-service attacks, Glob Denial-of-Service (\gdos), providing a framework for systematic improvement.
\item We propose and formalize \emph{GlobSpec}, a unified grammar and standard specification that establishes a strong foundation for glob pattern matching, as a path toward resolving the expressiveness and standardization gaps our study uncovered.
\end{itemize}

By addressing these gaps, our work lays the groundwork for more robust and portable globbing implementations, ultimately benefiting developers and improving software quality.

%% file: background.tex

Globbing\footnote{For the POSIX specification, see the GNU C Library manual on ``Pattern Matching'' (\url{https://ftp.gnu.org/old-gnu/Manuals/glibc-2.2.3/html_node/libc_137.html}). 
For detailed specifications, see the Linux \texttt{glob(7)} manual page (\url{https://man7.org/linux/man-pages/man7/glob.7.html})} refers to the process of expanding a pattern to 
match a set of pathnames. It originated in the early Unix shell with a simple, finite set of wildcards: $*$, $?$, and \texttt[abc]. This places globs squarely within the class of 
regular languages. However, globs represent a domain-specific language (DSL) specialized for hierarchical name structures in contrast to the general-purpose text matching of regex 
languages. Globbing simplicity is its strength, leading to widespread adoption far beyond the shell.

What began as simple command-line wildcard expansion has evolved into a ubiquitous mini-language with multiple dialects embedded across diverse computational contexts:
\begin{itemize}
\item \textbf{Source Control}: Git fundamentally relies on glob patterns in \texttt{.gitignore} files
\item \textbf{Build Systems}: Tools like Make and CI/CD pipelines employ globs for file specification and artifact management
\item \textbf{Web Development}: Patterns appear in URLs and URIs for hierarchical matching of routes and permissions in web services
\end{itemize}

However, this proliferation occurred without a unifying standard, resulting in a landscape of incompatible dialects. Implementations that vary in their support for essential 
features like recursive wildcards ($**$), negation ($!$), brace expansion ($\{\}$), and exhibit divergent behavior across platforms regarding case sensitivity, path separators, 
and hidden file handling. 

Globbing plays a crucial role in file system navigation and automation, enabling efficient pattern-based file retrieval. However, despite its widespread use, there is a notable 
lack of standardization across different programming languages, leading to inconsistencies in behavior, usability challenges, and security concerns~\cite{symbolic-execution-based-testing}. 
The variations in globbing behavior become particularly evident in cross-platform scenarios. Unix-like systems treat filenames starting with a dot (\texttt{.hidden}) as hidden 
files, which may be included or excluded in globbing results based on the implementation. In contrast, Windows does not follow the same convention, leading to differences when 
scripts are ported between platforms. Additionally, differences in case sensitivity (Unix is case-sensitive while Windows is not), and unicode handling further complicate cross-platform 
compatibility.

This proliferation without standardization has created an ecosystem where the same pattern may behave differently across contexts, leading to the reliability and security issues 
that this study empirically investigates. Given the widespread use of globbing in file handling and automation, it is crucial to understand these differences to avoid common 
pitfalls and ensure reliable, portable behaviors.

%% file: rqs.tex

In this work we seek to better understand globs as a feature, the developer practices around using them, and to support research targeting challenges that software developers face 
when using them. Thus, we focus on the following research questions:
\begin{enumerate}
\item [\textbf{RQ1}]Are features uniform across globbing implementations and, if not, how do these features differ?
\item [\textbf{RQ2}]Do differences in globbing behavior and dialect impact usability/reliability and to what extent? 
\item [\textbf{RQ3}]Are there common classes of correctness and security issues encountered with globs?
\item [\textbf{RQ4}]How comprehensive are globbing languages in addressing the tasks encountered in practice?
\end{enumerate}

%% file: methodology.tex

This study employs a mixed-methods approach to comprehensively investigate glob pattern usage across programming ecosystems. Our methodology integrates quantitative analysis of 
$62,713$ glob patterns extracted from $1,355$ open-source repositories with qualitative analysis of $2,160$ developer discussions from Stack Overflow, GitHub issues, and CVE 
databases. This multi-method approach enables us to establish both technical implementation landscapes and practical developer experiences.

\subsection{Language and Environment Selection}

We selected six environments that utilize glob patterns which represent diverse use-cases and implementation paradigms. Our primary selection criteria was covering the diverse set 
of scenarios where globbing has been integrated into the ecosystem as a DSL. This selection is based primarily in the authors $15$ years of experience working with large software 
companies, primarily Microsoft, and experience there with projects on build systems, programming language implementation, and cloud development. Secondarily, we used the TIOBE index 
to ensure we covered a diverse set of popular programming languages~\cite{tiobe}. The selected languages and their glob implementations are as follows: Bash (shell and file-system focused), 
JavaScript/Node.js (URI and server routes + git usage), Go (both standard library available via third-party extensions), PHP and Ruby (legacy web sites and contrasting 
builtin vs library support), and Python (as a general purpose language). 

\begin{itemize}
    \item \textbf{Bash} was selected due to its foundational role in Unix-like operating systems and its extensive use in scripting and automation tasks that heavily interact with file 
    system style operations.
    \item \textbf{JavaScript/Node.js} was chosen for its ubiquity in modern web development, where glob patterns are frequently used for file management, build processes, and routing in 
    frameworks like \emph{Express.js}.
    \item \textbf{PHP and Ruby} were included to represent server-side web development environments where glob patterns are often utilized in file handling and asset management tasks. These 
    languages also provide a contrast between built-in globbing capabilities (Ruby) and third-party library implementations (PHP).
    \item \textbf{Go} was selected for its growing popularity in systems programming and cloud-native applications, where efficient file pattern matching is crucial. Go's standard library includes builtin 
    globbing functionality, and is complemented by several third-party libraries that extend its capabilities.
    \item \textbf{Python} was included due to its versatility and widespread adoption across various domains, including data science, web development, and automation. Globbing in Python 
    is also implemented in multiple, slightly different ways, including the built-in \texttt{glob} module and \texttt{pathlibPath.glob()}.
\end{itemize}

For each language, we analyzed the canonical (or popular $3^{\text{rd}}$ party) glob implementation: Python's \texttt{glob} and \texttt{pathlib.Path.glob()}, 
Node.js's \texttt{glob} package (200+ million weekly downloads), Ruby's \texttt{Dir.glob()}, Go's \texttt{filepath.Glob()} with third-party alternatives, PHP's \texttt{glob()} 
function, and Bash's native expansion. We conducted cross-platform testing on Linux, macOS, and Windows to evaluate semantic differences in case sensitivity, path separator handling, 
and hidden file behavior.

\subsection{Repository Selection and Pattern Extraction}

We constructed a representative corpus through stratified sampling of $1,966$ active GitHub repositories (Python: $377$, JavaScript: $365$, Shell: $331$, Go: $314$, PHP: $300$, Ruby: $279$) 
meeting these criteria:

\begin{itemize}
\item Primary language composition $>50\%$ to ensure idiomatic usage
\item Minimum 150 GitHub stars indicating community adoption
\item Active maintenance (commits within preceding 12 months)
\item Exclusion of archived or deprecated repositories
\end{itemize}

We utilized a suite of per-language regular expressions, included in the reproduction package, for extracting and analyzing glob usage. Post extraction we manually reviewed a random 
sample of $150$ repositories to evaluate extraction accuracy. This validation showed a recall rate of $94\%$ and a precision rate of $99\%$.

\begin{figure}[htbp]
    \centering
    \includegraphics[width=\columnwidth]{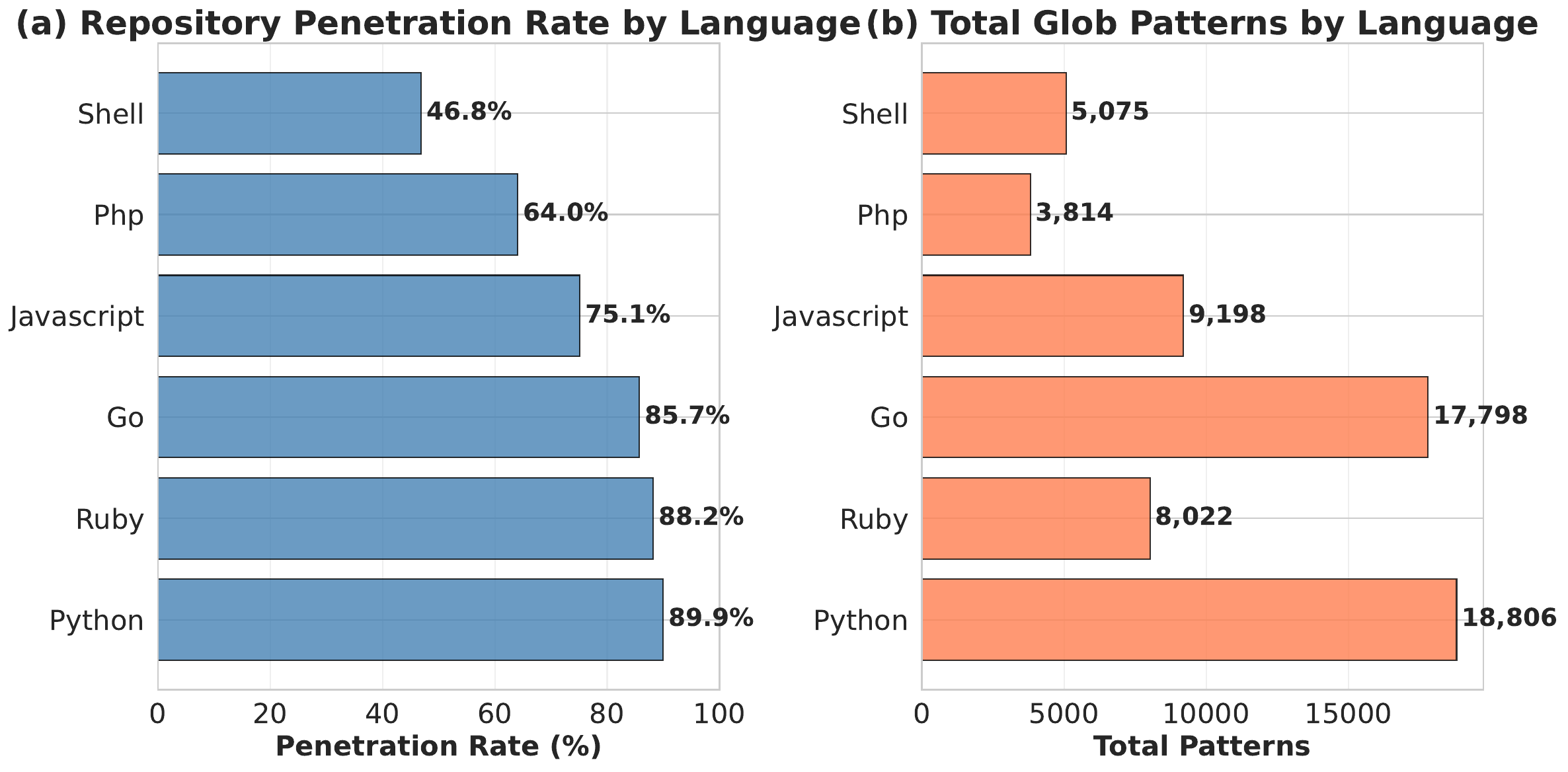}
    \caption{Distribution of glob pattern occurrence across programming languages. Fig. (a) shows the percentage of repositories containing at least one glob pattern. Fig. (b) 
    reports the total number of globs extracted from repositories in each language.}
    \label{fig:language_prevalence}
\end{figure}

\subsection{Glob Frequency and Distribution}

Our language selection strategy targeted six programming ecosystems representing diverse implementation approaches and application domains. 
\Cref{fig:language_prevalence} shows significant variation in both repository penetration (the percentage of repositories containing glob patterns) 
and total pattern volume across languages. Python demonstrates the highest adoption with a number of $18,806$ of all extracted patterns, reflecting 
its prevalence in file-system-intensive domains like data science and scripting. Go follows with $17,798$, while Node.js, PHP, Ruby, and Shell show 
more modest but strategically important usage patterns. This distribution informed our stratified sampling approach, ensuring adequate representation 
of both high-volume ecosystems and specialized use cases while maintaining statistical power for cross-language comparison.

\begin{figure}[htbp]
    \centering
    \includegraphics[width=\columnwidth]{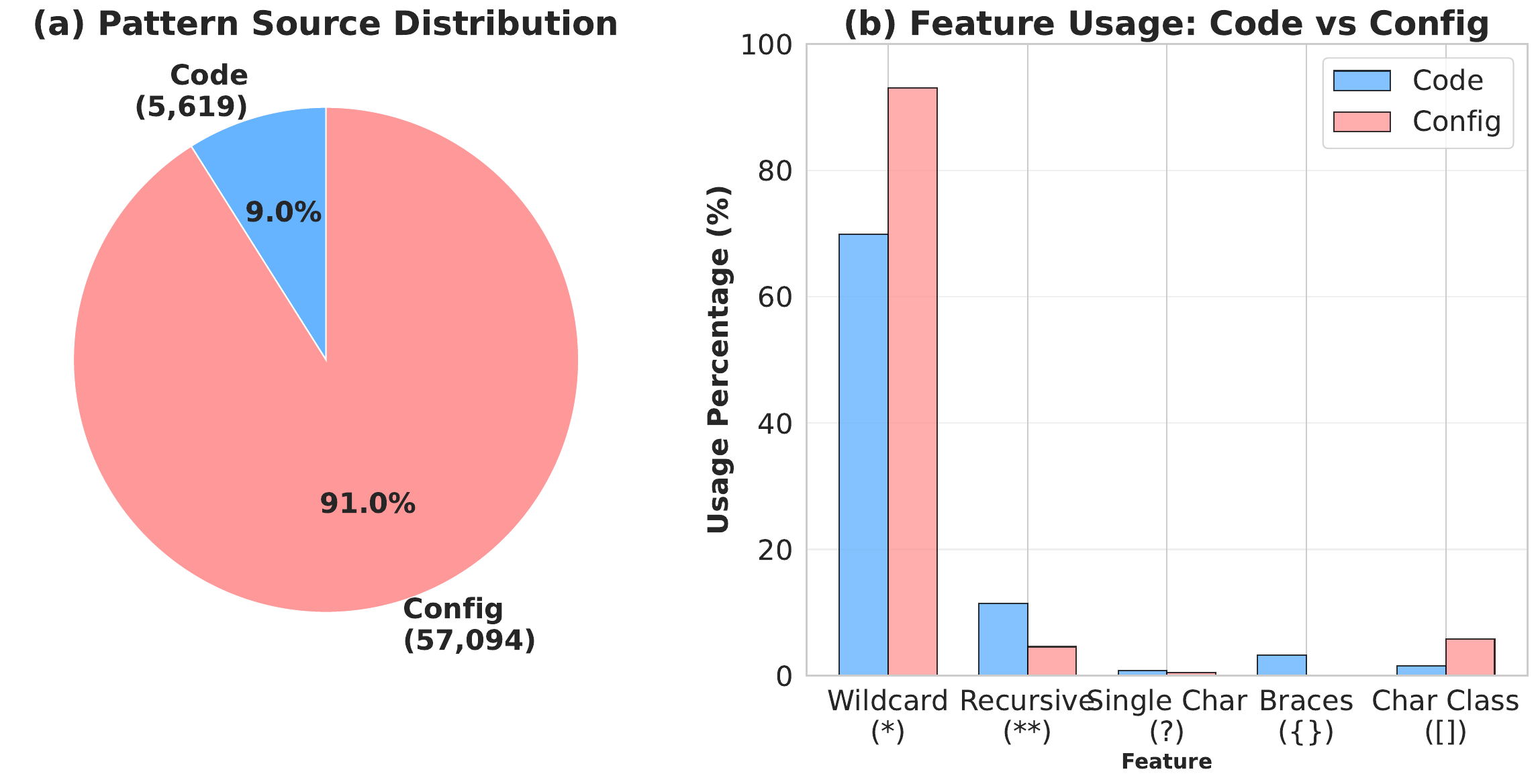}
    \caption{Distribution of glob patterns by source file type, comparing configuration files (YAML, JSON, INI configs) versus language source code files across all studied languages. 
    Across all languages The vast majority of glob patterns are found in configuration files.}
    \label{fig:source_comparison}
\end{figure}

\Cref{fig:language_prevalence} (b) shows the final corpus comprises $62,713$ validated patterns with distribution: Python $(29.99\%)$, JavaScript $(14.67\%)$, 
Go $(28.38\%)$, PHP $(6.08\%)$, Ruby $(12.79\%)$, Shell $(8.09\%)$. \Cref{fig:source_comparison} shows configuration files $(91.0\%)$ 
dominate source code use $(9.0\%)$. This bias towards configuration contexts, along with the prevalent use of glob patterns in code to describe resources such as 
file paths or URL routes, is notable and suggests that glob patterns are primarily employed for declarative resource specification rather that for search/replacement 
style tasks that are seen with regular expressions.




\subsection{Qualitative Data Collection and Analysis}

Our qualitative data study utilizes $2,160$ developer discussions from three complementary sources ranging in time from 2018-2024 to capture a broad spectrum of glob-related issues 
involving usability, reliability, security, and performance concerns:

\begin{itemize}
\item \textbf{Stack Overflow}: $361$ questions revealing help-seeking behavior and conceptual difficulties
\item \textbf{GitHub Issues}: $1,355$ reports documenting production problems and feature requests  
\item \textbf{CVE Database}: $444$ entries cataloging security vulnerabilities and exploits
\end{itemize}

We performed API scraping with keyword-based filtering -- (core: "glob", "wildcard"; feature-specific: "recursive glob", "brace expansion"; problem-specific: "glob error", "security")~\cite{labelprediction}
to collect relevant posts. We excluded low-score discussion posts and fewer interations Github issues. The resulting corpus was manually reviewed to ensure topical relevance, yielding a final dataset of $2,160$ posts 
which are included in the reproduction artifact.

Using an iterative hybrid coding approach, combining deductive and inductive methods outlined below, we developed a taxonomy of glob-related developer issues. 

\begin{enumerate}
\item \textbf{Deductive initial categories}: Derived from software engineering literature: Usability, Reliability, Security, Performance, Feature Gaps, Compatibility
\item \textbf{Automated categorization}: Keyword-based scoring assigned posts to primary categories
\item \textbf{Inductive subcategory refinement}: Phrase frequency analysis identified recurring themes within each category
\item \textbf{Validation}: Analysis of "Other" categories confirmed taxonomy completeness despite natural heterogeneity in developer problem descriptions
\end{enumerate}

In addition to developer initiated discussion we also conducted an systematic review of the official documentation for all studied glob 
implementations~\cite{BashGlob,GoGlob,RubyGlob,PythonFnmatch,PythonGlob,PythonPathlib,PHPGlob,NodeGlob}. This analysis identified documentation gaps, platform-specific 
behavior omissions, and evolutionary changes across library versions. For instance, we tracked Python's Unicode handling improvements from version $2$ to 
$3$~\cite{python2-3}, Node.js's recursive matching enhancements, and Bash's extended glob support in version $4.0+$.

%% file: analysis.tex
Based on the data collection and methodology described in \Cref{sec:methodology}, we present our analysis addressing each research question in turn.

\subsection*{\bf RQ1: Are features uniform across globbing implementations and, if not, how do these features differ?}

\begin{figure}[ht]
    \centering
    \includegraphics[width=\columnwidth]{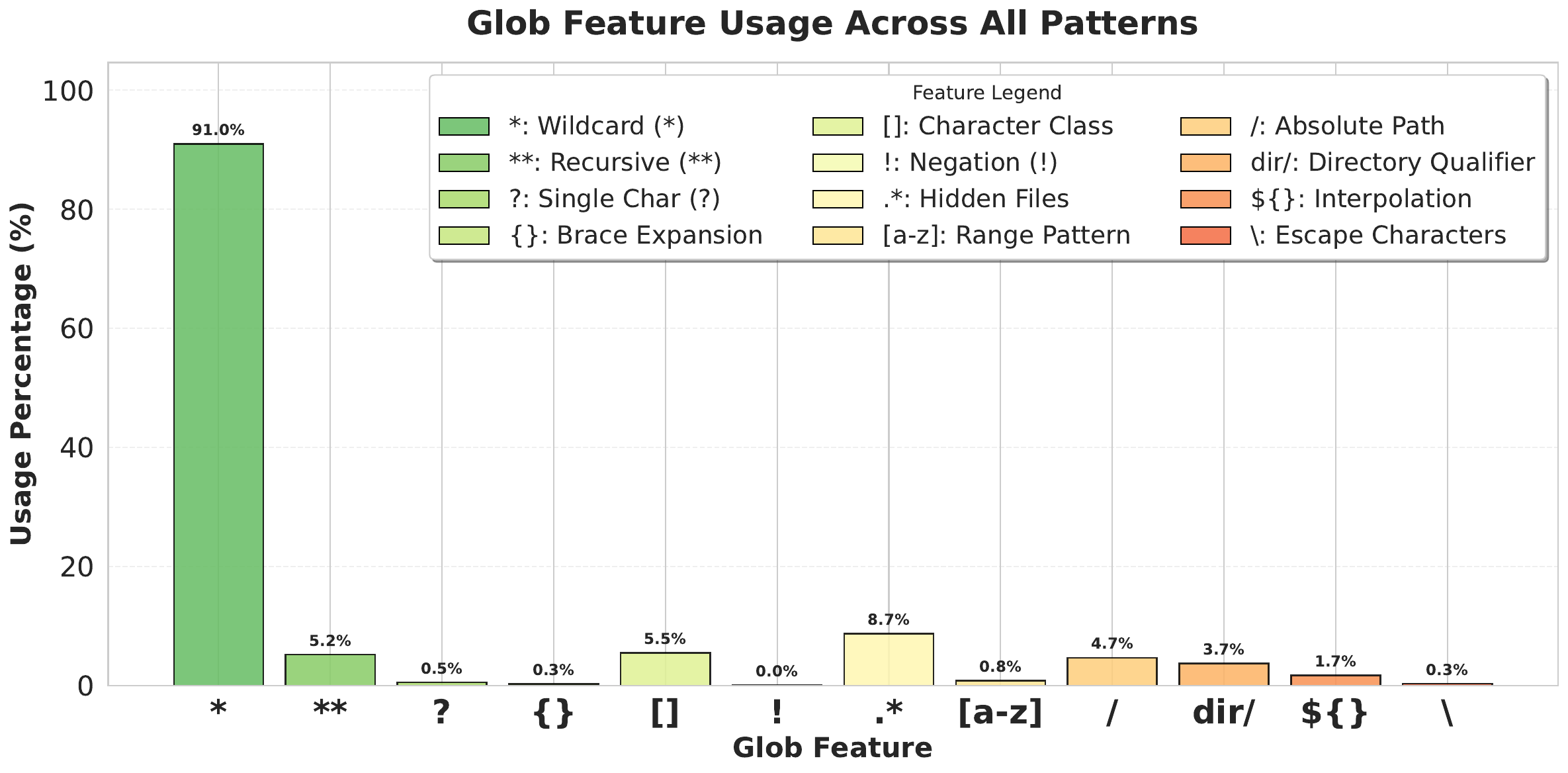}
    \caption{Empirical usage patterns of glob features across $62,713$ patterns extracted from open-source repositories}
    \label{fig:feature_usage}
\end{figure}

\Cref{tab:glob_features_support} and \Cref{fig:feature_usage} reveal significant disparities between glob feature availability and actual usage patterns across programming languages.
\begin{table}[t]
    \centering
    \small
    \caption{Support for Glob features across languages based on official language or $3^{\text{rd}}$ party library documentation.}
    \label{tab:glob_features_support}
    \begin{tabular}{lcccccc}
        \toprule
        \textbf{Feature} & \textbf{Bash} & \textbf{Python} & \textbf{Node.js} & \textbf{Ruby} & \textbf{PHP} & \textbf{Go} \\
        \midrule
        Wildcard (*)         & $\bullet$ & $\bullet$ & $\bullet$ & $\bullet$ & $\bullet$ & $\bullet$ \\
        Recursive (**)       & $\bullet$ & $\bullet$\textsuperscript{1} & $\bullet$ & $\bullet$ & $\circ$ & $\circ$ \\
        Single-char (?)      & $\bullet$ & $\bullet$ & $\bullet$ & $\bullet$ & $\bullet$ & $\bullet$ \\
        Brace expansion (\{\}) & $\circ$ & $\bullet$ & $\bullet$ & $\bullet$ & $\bullet$ & $\bullet$\textsuperscript{2} \\
        Character class ([]) & $\bullet$ & $\bullet$ & $\bullet$ & $\bullet$ & $\bullet$ & $\bullet$ \\
        Negation (!)         & $\circ$ & $\circ$ & $\bullet$ & $\circ$ & $\bullet$ & $\circ$ \\
        Hidden files (.*)    & $\bullet$ & $\bullet$\textsuperscript{3} & $\bullet$\textsuperscript{4} & $\bullet$\textsuperscript{5} & $\bullet$ & $\bullet$ \\
        \bottomrule
    \end{tabular}
    
    \vspace{2mm}
    \small
    \raggedright
    \textbf{Legend}: $\bullet$ = Native support; $\circ$ = No native support. \\
    \textsuperscript{1}\texttt{recursive=True} required,
    \textsuperscript{2}\texttt{GLOB\_BRACE} flag required,
    \textsuperscript{3}\texttt{include\_hidden=True} required,
    \textsuperscript{4}\texttt{dot:true} option required,
    \textsuperscript{5}\texttt{File::FNM\_DOTMATCH} required
\end{table}

The data in \Cref{tab:glob_features_support} shows that four features have universal support: wildcard (\texttt{*}), single-character (\texttt{?}), character classes (\texttt{[]}), and 
hidden file (\texttt{.*}). These form a minimal portable core derived from historical Unix implementations. 

An analysis of the $62,713$ glob patterns reveals stark usage disparities. \Cref{fig:feature_usage} demonstrates clear hierarchy of feature adoption. The wildcard operator (\texttt{*}) 
dominates with $90.98\%$ prevalence ($57,061$ patterns), confirming its role as the fundamental glob construct. Hidden file patterns appear in $8.72\%$ of cases ($5,470$ patterns), 
reflecting the importance of configuration file management. Character classes ($5.48\%$, $3,437$ patterns) and recursive globbing ($5.21\%$, $3,265$ patterns) show moderate adoption 
for specific matching needs. Most strikingly, many other features exhibit minimal usage: single-character wildcards ($0.5\%$, $322$ patterns), brace expansion ($0.29\%$, $182$ patterns), 
and negation ($0.02\%$, $11$ patterns). 

This distribution reveals a significant gap between available language features and actual developer adoption, suggesting that current implementations either fail to meet practical 
needs or present usability barriers for advanced functionality. Beyond the core features we immediately encounter problematic fragmentation in behaviors and feature availability including:

\begin{itemize}
\item \textbf{Recursive globbing (\texttt{**})}: Python requires explicit flags, Go lacks native support, Bash needs optional configuration.
\item \textbf{Brace expansion}: is implemented at different abstraction layers (shell vs. library) and result in behavioral variations.
\item \textbf{Negation}: is implemented via multiple incompatible mechanisms (gitignore prefix, character class, extended glob).
\item \textbf{Hidden files}: have inconsistent defaults, creating security risks across APIs, even within the same language.
\end{itemize}

\subsection*{\bf RQ2: Do differences in globbing behavior and dialect impact usability/reliability and to what extent? }
\setcounter{subsection}{0}

\begin{figure}[ht]
    \centering
    \includegraphics[width=\columnwidth]{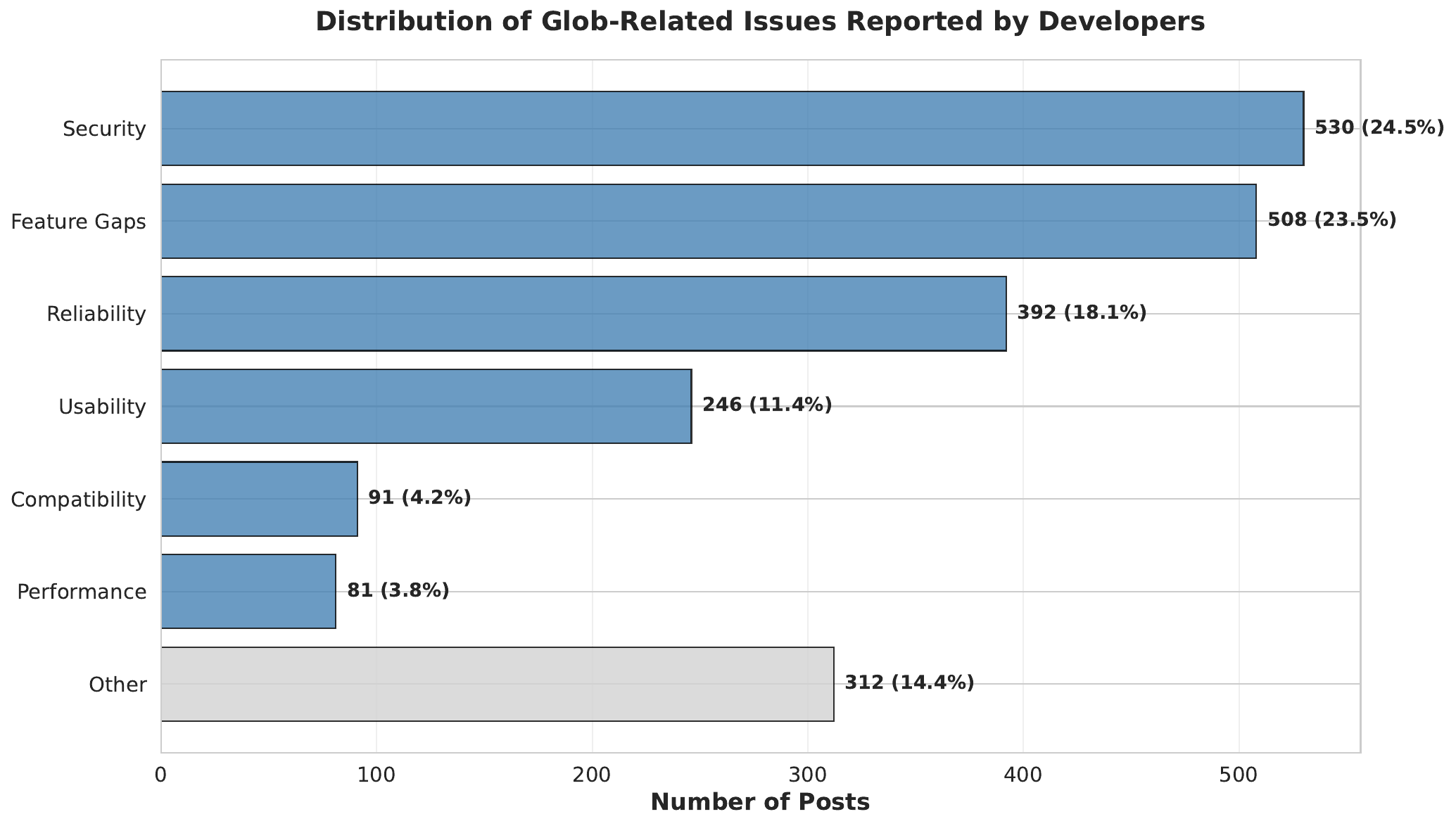}
    \caption{Distribution of $2,160$ glob-related issues, derived from developer discussions, across six primary categories -- security, features, reliability, usability, 
    compatibility, performance, and other.}
    \label{fig:issue_categories}
\end{figure}

The analysis of developer discussions ($2,160$ in total) reveals how glob implementation inconsistencies translate into 
reliability and usability problems, with platform-specific semantic drift emerging as a critical concern. We enumerate the general themes identified in our taxonomy below 
(\Cref{fig:issue_categories}) and provide more detailed analysis and case studies for the most significant issues.

Glob issues span a range of severity levels, with critical problems comprising $22.3\%$ of reports. The taxonomy distribution (\Cref{fig:issue_categories}) shows reliability: 
$392$ posts ($18.1\%$), usability: $246$ posts ($11.4\%$), feature gaps: $508$ posts ($23.5\%$), security: $530$ posts ($24.5\%$), compatibility: $91$ posts ($4.2\%$), 
performance: $81$ posts ($3.8\%$).

Language-specific analysis reveals disproportionate friction in certain ecosystems. Go appears in $523$ issue mentions ($24.2\%$ of language references) despite representing 
only $28.38\%$ of our pattern corpus, indicating significant challenges with its standard library's limited glob support and third-party fragmentation.

Analysis of subcategories within the reliability issues shown in \Cref{fig:issue_categories} reveals that runtime errors dominate reliability issues ($190$ posts, $48.5\%$). 
The high proportion of runtime errors suggests that glob implementations often encounter conditions that were not expected by developers, leading to application crashes or unhandled exceptions during file system operations. This pattern indicates potential gaps in error 
handling robustness across different glob implementations.

The related issue of cross-tool inconsistency ($40$ posts, $10.2\%$) demonstrates how patterns working in one tool fail in another. These cases reveal that developers expect 
consistent glob behavior across different tools and libraries, but encounter unexpected differences in pattern matching semantics. The presence of these inconsistencies 
suggests the lack of standardized behavior for glob implementations causes substantial friction, forcing developers to learn and account for tool-specific variations in their workflows, 
and inhibits the use of globbing.

Integration problems ($32$ posts, $8.2\%$) occur when glob patterns work in isolation but fail in system contexts. These issues highlight how glob operations are sensitive to 
their runtime environment, including factors such as current working directory, user permissions, and filesystem state. The recurrence of integration problems suggests that glob 
patterns may have implicit dependencies on execution context that are not always apparent during development or testing in isolated environments.

Within the usability issues shown in \Cref{fig:issue_categories}, documentation gaps represent a significant usability barrier, comprising $64$ posts ($26.1\%$ of usability issues). Developers consistently report insufficient documentation 
covering edge cases, platform-specific behaviors, and advanced feature usage. 

Syntax confusion and error messages, appearing in $18$ posts ($7.3\%$) and $7$ posts ($2.9\%$) respectively, involve mismatches between developer expectations from shell 
experience and programmatic glob behavior along with generic error messages making issues difficult to investigate. 

\subsection{Hidden File Handling Complications}
Despite $5,470$ patterns ($8.72\%$) explicitly referencing hidden files, inconsistent defaults create persistent hazards and information disclosure risks across programming ecosystems. 
The problem is particularly acute when recursive globbing interacts with hidden file policies, as demonstrated by Node.js's inconsistent behavior:

\begin{lstlisting}[language=JavaScript, basicstyle=\ttfamily\footnotesize]
// These patterns match hidden files:
'subdir/{.,}*.js'    // Explicit directory
'*/{.,}*.js'         // Single wildcard

// These patterns unexpectedly exclude hidden files:
'scan/**/{.,}*.js'   // Double wildcard excludes hidden files
'**/{.,}*.js'        // Also excludes hidden files
\end{lstlisting}

Inconsistency across implementations results in the same pattern producing different results across tools. While \texttt{node-glob} and bash correctly match hidden files with recursive 
wildcards, Node.js's built-in \texttt{fs/promises} glob implementation unexpectedly excludes them. This creates silent failures where deployment scripts work during development with 
one library but fail (or silently produce different results) when passed to another. For example, a deployment script using \texttt{dist/**/*} will match files inside the hidden directory \texttt{dist/.cache} 
when using node-glob, but silently exclude them when the same pattern is used with Node.js's \texttt{fs/promises} glob. This potentially breaks the deployed application if it depends on 
cached assets.

Security and reliability are major concerns with the inconsistent behavior around hidden files which often contain important configs information (which should be included) or sensitive 
forms of data (which must be excluded). Build processes may miss critical configuration files, causing runtime failures. Conversely, when hidden files are unexpectedly included, deployment 
scripts may expose sensitive files like \texttt{.env} or \texttt{.credentials} containing API keys and database passwords. This vulnerability is documented in CVE-2021-42135, where glob pattern
ambiguity in HashiCorp Vault's policy system allowed unauthorized access to credential generation endpoints.

The Python ecosystem is another example with divergent hidden path behavior for \texttt{glob.glob()} (excludes hidden files by default) and \texttt{pathlib.Path.glob()} (includes them). 
These inconsistencies cause production incidents where CI environment behavior diverges from development setups, creating unreliable deployments and potential information 
disclosure vulnerabilities.

\subsection{Platform-Specific Semantic Drift}
Our data analysis identified $91$ posts ($4.2\%$ of total corpus) explicitly describing compatibility problems, with platform differences dominating ($49$ posts, $53.85\%$ of compatibility 
issues). 

Case sensitivity mismatches create silent failures when moving between Unix-like systems and Windows. A Stack Overflow question illustrates: "I have files named \texttt{1\_1.txt}, 
\texttt{1\_2.TXT}, \texttt{2\_1.txt}. On Windows, \texttt{glob.glob('*.txt')} matches all three files. On Linux, it only matches lowercase extensions. My code broke because it depends 
on processing all files regardless of capitalization."


Unicode normalization differences are a cause of failures that are particularly difficult to diagnose. File systems can differ in normalization schemes (NFC vs. NFD), causing the same 
visual character to be stored using different byte sequences.

\begin{lstlisting}[language=Python, basicstyle=\ttfamily\footnotesize]
# Matches NFD-encoded 'café' but may miss NFC 'café'
files_macos = glob.glob('*café*')    
# Behavior varies by filesystem: may match NFC, NFD, or both
files_linux = glob.glob('*café*')    

# Underlying encoding differences:
# macOS (HFS+/APFS): NFD - 'cafe' + combining acute
# Windows/Linux: NFC - precomposed 'é' character
\end{lstlisting}

As noted in developer forums: "macOS stores file names in Unicode NFD, while other OSes like Windows use NFC. Linux filesystems can use both encodings." This creates subtle cross-platform 
bugs with deployment scripts working during testing on macOS but breaking in production on Linux servers. The problem may also manifest as missing files or duplicate files when sharing 
across systems, requiring manual normalization workarounds that current glob implementations lack.

Locale-dependent character classes amplify these issues and create additional platform variation. The pattern \texttt{[a-z]} behavior depends on system locale settings: under UTF-8 locales, 
ranges may match letters with diacritics, while C locale (ASCII-only) excludes them. One Bash-focused post describes production incidents where \texttt{rm [a-z]*} deleted files like 
\texttt{élite.txt} that weren't removed during development.

Finally, version incompatibility ($7$ posts, $7.69\%$ of compatibility issues), describes breaking changes between library versions. A representative case documents Python $3.8$ to $3.10$ 
migration where \texttt{glob.glob()} changed default symlink following behavior for security reasons, causing production behavior to differ from development.

The platform-specific semantic drift particularly affects cross-platform development workflows \cite{lingua-franca}, where subtle behavioral differences create bugs that are difficult 
to reproduce and diagnose. These issues often manifest only in production environments, making them expensive to discover and resolve.

\subsection*{\bf RQ3: Are there common classes of correctness and security issues encountered with globs?}
\setcounter{subsection}{0}

Our analysis identified $530$ security-related posts from the discourse corpus ($24.5\%$ of total). These posts reveal that, in addition to theoretical concerns, glob security problems 
manifest as actual exploited vulnerabilities as well. \Cref{fig:issue_categories} shows that security issues represent the largest category of developer concerns, underscoring the 
critical importance of addressing glob-related vulnerabilities in production systems.

\begin{figure}[ht]
    \centering
    \includegraphics[width=\columnwidth]{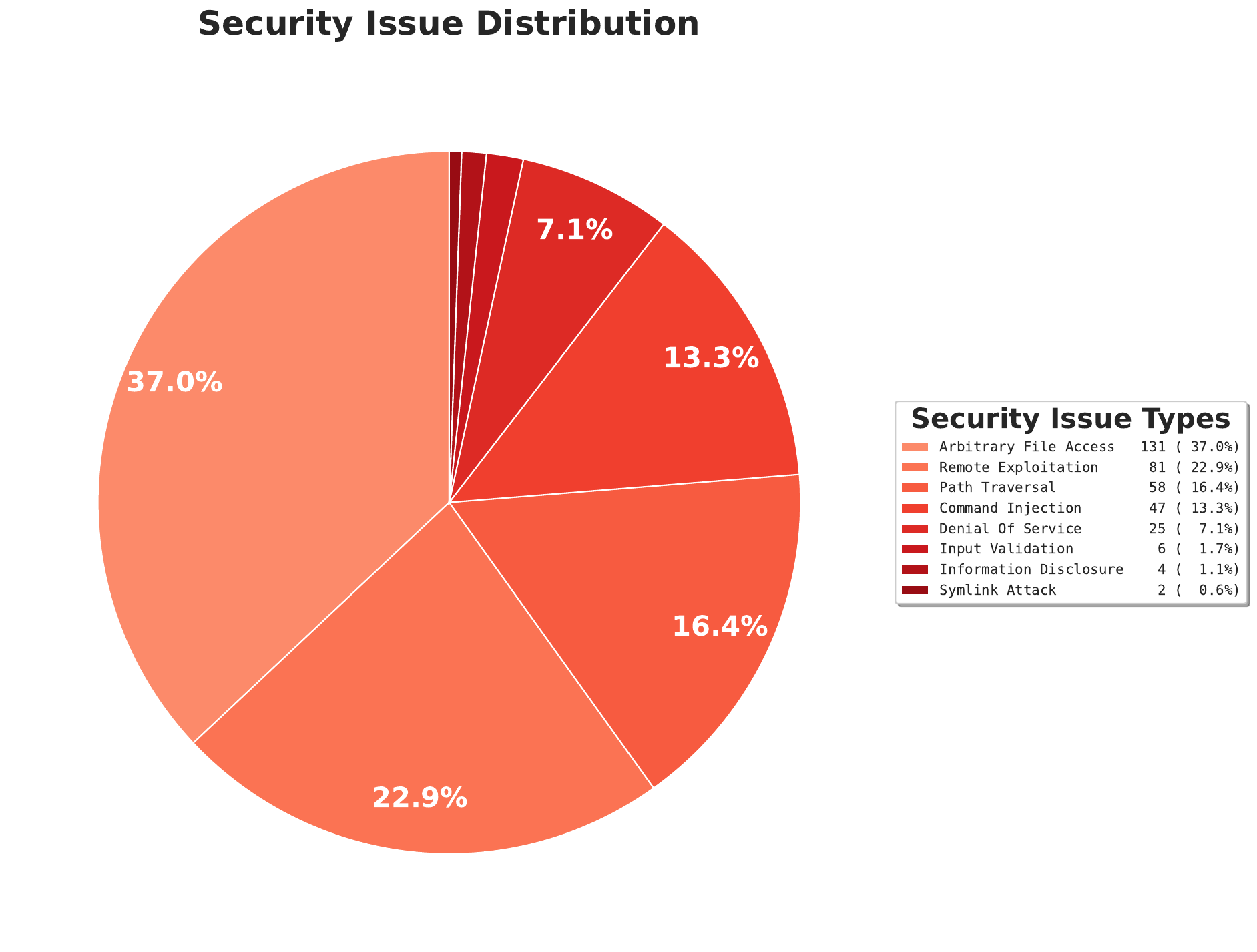}
    \caption{Detailed breakdown of 528 security issues based on analyzed developer discussions, highlighting prevalent vulnerability classes and showing the contribution of glob based 
    denial-of-service attacks, which represent a emerging threat vector globs are increasingly used in validation checks for URI resource access and API path routing.}
    \label{fig:security_taxonomy}
\end{figure}

Our study reveals a nuanced threat landscape dominated by unauthorized file access, with detailed breakdown shown in \Cref{fig:security_taxonomy}. The security taxonomy, derived from 
analysis of our corpus of dataset, demonstrates that glob vulnerabilities extend beyond simple path traversal to encompass sophisticated attack vectors.

The distinction between Arbitrary File Access ($37.8\%$) and Path Traversal ($16.4\%$) is crucial: while Path Traversal specifically involves directory escape sequences, Arbitrary 
File Access encompasses broader exploitation patterns including symlink attacks, predictable filename attacks, and glob pattern injection that construct valid paths to sensitive files.

\subsection{Glob-based Denial-of-Service (G-DoS) Vulnerabilities}

We identify \textbf{\gdos} as a distinct denial-of-service vector characterized by resource exhaustion through pathological pattern matching (similar to the rise of ReDoS~\cite{impact-of-ReDoS,ReDoS-web} 
attacks). DoS vulnerabilities~\cite{craft-ReDoS,ReDoS-web,static-detection-dos} can be used to both attack the glob matching engine itself~\cite{rethink-regex} but 
can also be performed by overwhelming the underlying file system (or resource) through excessive I/O operations.

Vulnerability CVE-2022-30630 existed in Go's original \texttt{Glob} function, which invoked recursive logic without depth bounds:

\begin{lstlisting}[language=Go, basicstyle=\ttfamily\footnotesize
]
func Glob(pattern string) (matches []string, err error) {
    if _, err := Match(pattern, ""); err != nil {
        return nil, err
    }
    if !hasMeta(pattern) {
        if _, err = os.Lstat(pattern); err != nil {
            return nil, nil
        }
        return []string{pattern}, nil
    }
    // ... recursive logic without depth limits
}
\end{lstlisting}

The fix introduced a hard-coded depth limit:

\begin{lstlisting}[language=Go, basicstyle=\ttfamily\footnotesize
]
func Glob(pattern string) (matches []string, err error) {
    return globWithLimit(pattern, 0)
}

func globWithLimit(pattern string, depth int) (matches []string, err error) {
    const pathSeparatorsLimit = 10000
    if depth == pathSeparatorsLimit {
        return nil, ErrBadPattern
    }
    // ... original logic
}
\end{lstlisting}

While preventing crashes, this solution is brittle and the core issue remains. The arbitrary 10,000 separator limit creates new failures: legitimate operations on deeply 
nested directories (e.g., \texttt{node\_modules}) are abruptly terminated. This transforms potential DoS attacks into guaranteed failures for legitimate users exceeding the 
threshold \cite{Golangvuln}. An alternative attack~\cite{heapoutofbound} highlights heap out-of-bounds issues, attacking the engine itself, triggered through recursive globbing. 
The Apple platform vulnerabilities~\cite{apple1,apple2,apple3} demonstrate how crafted strings can cause excessive resource consumption in \texttt{libc} glob implementations.

\subsection{Information Disclosure Vulnerabilities}

Information disclosure vulnerabilities ($4$ posts, $1.1\%$) have a disproportionately severe impact despite low frequency~\cite{dotfilesecurity}, particularly through 
inadvertent exposure of secret data in configuration files. The most common scenario involves patterns like \texttt{public/**/*} intended to match public assets but inadvertently 
including configuration files which may contain keys or other secrets. The HashiCorp Vault vulnerability (CVE-2021-42135) exemplifies how semantic ambiguity in glob 
patterns can directly compromise security boundaries.

\begin{lstlisting}[language=HCL, basicstyle=\ttfamily\footnotesize]
path "gcp/roleset/*" {
  capabilities = ["read"]
}
# In Vault 1.8.0+, this policy unintentionally granted permission to generate Google Cloud service account credentials
\end{lstlisting}

Prior to Vault 1.8.0, the pattern \texttt{gcp/roleset/*} correctly provided read-only access to roleset listings. The introduction of roleset-specific credential endpoints at 
paths like \texttt{gcp/roleset/\{name\}/credential} caused the existing glob pattern to match these new, security-critical paths. The \texttt{read} capability, which was 
safe for enumeration operations, became dangerous when applied to credential generation endpoints. This semantic misinterpretation violated the principle of least privilege 
and allowed attackers to generate arbitrary Google Cloud service account credentials. This example illustrates how a glob vulnerability was leveraged to create a privilege 
escalation vector for full cloud environment compromise.

\subsection{Unexpected Matching and Pattern Ambiguity}

Unexpected matching represents a significant correctness concern, particularly with hidden file patterns. The \texttt{.*} pattern, while intended to match all hidden files in 
a directory, also matches the current directory (\texttt{.}) and parent directory (\texttt{..}), leading to incorrect results and potential errors in file processing.
The lack of consistent hidden file handling across programming languages exacerbates these issues, forcing developers to implement additional filtering logic. Patterns that 
work correctly in one environment may produce unexpected matches in another due to different default behaviors for special directories and hidden files.
Case sensitivity inconsistencies introduce additional unexpected matching risks. As demonstrated by CVE~\cite{caseinsensitive}, case-insensitive pattern matching can result in 
unauthorized file access when moving between platforms with different filesystem semantics.

\subsection{Path Traversal Vulnerabilities}

Path traversal remains a critical security concern, with $58$ posts ($16.4\%$ of security issues) specifically involving directory escape sequences. According to the MITRE CWE Top 
25 Most Dangerous Software Weaknesses of 2024~\cite{mitre2024top25}, path traversal (CWE-22) ranked 5th, underscoring its prevalence and impact.
Poorly constructed glob patterns such as \texttt{../../*} allow attackers to traverse directory hierarchies and access sensitive system files. This issue is particularly concerning 
in web applications, command-line utilities, and API endpoints.
The vulnerability~\cite{globinconsistency} exposes how inconsistencies in glob handling can be exploited to bypass intended file restrictions. Similarly,~\cite{insecureparsing} 
and~\cite{pathescape} demonstrate how improper escaping and parsing of special characters lead to unintended behavior and security vulnerabilities.

\subsection*{\bf RQ4: How comprehensive are globbing languages in addressing the tasks encountered in practice?}
\setcounter{subsection}{0}

Our analysis reveals significant gaps between developer needs and current glob implementations across documentation, feature completeness, usability, and performance. 
By synthesizing findings from all research questions, we identify strategic improvement priorities grounded in empirical evidence from $2,160$ developer discussions 
and $62,713$ pattern analyses.

\subsection{Documentation Inadequacy}

Documentation gaps represent the most significant usability barrier, comprising $63$ posts ($25.7\%$ of usability issues). Developers consistently report insufficient coverage of edge cases, platform-specific behaviors, and practical usage scenarios. These documentation deficiencies directly contribute to the reliability and security issues identified in RQ2 and RQ3, as developers operate with incomplete mental models of glob behavior.

\subsection{Feature Gaps: Critical Missing Capabilities}

Feature gap complaints ($508$ posts) reveal critical deficiencies in current implementations, particularly around negation and efficient exclusion. Missing Features dominate with 
$100$ posts ($19.7\%$), indicating fundamental gaps in glob functionality. Developers frequently request capabilities like directory 
exclusion during traversal, case-insensitive matching, and result streaming interfaces.

The Negation Paradox is particularly striking: negation is among the most requested features ($150$ posts, $29.5\%$ of feature gaps). It is virtually absent in practice, appearing 
in only ($0.02\%$) of our $62,713$ extracted patterns. This stark contrast between demand and usage points to a critical failure in current implementations, where available negation 
features are likely difficult to discover, unreliable, or flawed. 

\begin{lstlisting}[language=JavaScript, basicstyle=\ttfamily\footnotesize]
// Works as expected:
glob('**/*', {nodir: true, ignore: '**/*.css'}, cb)

// Actual problematic behavior:
glob('./**/*', {nodir: true, ignore: '**/*.css'}, cb)
// All files matching the glob patterns are passed, 
// including files fulfilling the ignored glob
\end{lstlisting}

For example the npm Glob library provides negation {\bf but} it is done via a configuration option on an optional \cf{config} object and counterintuitively named \cf{ignore}. Further, 
the implementation of the features does not skip ignored paths during enumeration but, instead, fully enumerates all paths as specified by the glob, and then post filters out based 
on the specified ignore glob. Thus, while present, the feature is difficult to discover and inefficient to use.

Brace Expansion, as a Niche Demand, comprises $13$ posts ($2.6\%$), showing extremely low usage ($0.29\%$, $182$ patterns) yet is frequently requested by shell-experienced developers. 
This suggests brace expansion serves specialized workflows where its absence is blocking. Given the bias in requester expertise, this feature may also be universally useful but 
conceptually undiscovered by less experienced users.

\subsection{Usability Challenges}

Syntax Confusion appears in $13$ posts ($5.3\%$), involving mismatches between shell experience and programmatic behavior. 

\begin{lstlisting}[language=Python, basicstyle=\ttfamily\footnotesize]
from pathlib import Path
import glob
# Directory structure with symlink:
# C:\Folder
# C:\Folder\Subfolder -> D:\Subfolder (symlink)
# D:\Subfolder\File.txt

pathlib_result = list(Path('C:/Folder').glob('**/*'))
# Returns: [WindowsPath('C:/Folder/Subfolder')] - doesn't follow symlink

glob_result = glob.glob('C:/Folder/**/*', recursive=True)  
# Returns: ['C:/Folder/Subfolder', 'C:/Folder/Subfolder/File.txt'] 
\end{lstlisting}

This inconsistency, documented in Python/cpython issues $\#77609$, demonstrates how the same glob pattern yields fundamentally different results depending on the implementation 
used. The divergence creates reliability risks where scripts may miss critical files or unexpectedly include external directory contents.

Unexpected Behavior accounts for $7$ posts ($2.9\%$), where glob operations produce results that contradict developer expectations. A striking example emerges from 
Python's standard library, where the recursive wildcard pattern \texttt{**} exhibits three different behaviors across modules.

\begin{lstlisting}[language=Python, basicstyle=\ttfamily\footnotesize]
# Directory structure:
# toplevel.txt
# dir/
#   indir.txt
#   subdir/

import glob, pathlib

# Three conflicting interpretations of '**':
glob.glob('**')
# ['dir', 'toplevel.txt'] - files and dirs, non-recursive, not including .

glob.glob('**', recursive=True)  
# ['dir', 'dir/subdir', 'dir/indir.txt', 'toplevel.txt'] 
# - files and dirs, recursive, not including .

list(pathlib.Path('.').glob('**'))
# [PosixPath('.'), PosixPath('dir'), PosixPath('dir/subdir')]
# - only directories, recursive, includes .
\end{lstlisting}

As documented in Python issue \#106747, "the Python docs for pathlib.glob are unclear on the meaning of \texttt{**}" and the behavior "gets even more confusing" across implementations. 
The critical inconsistency lies in hidden directory handling: while both \texttt{glob.glob} variants explicitly exclude the current directory (\texttt{.}), the implementation in \texttt{pathlib.Path.glob} 
unexpectedly includes it. This creates subtle bugs when scripts attempt file operations on \texttt{.}, treating it as a regular file path.

This fragmentation demonstrates how semantic ambiguity in core patterns creates unexpected behavior within a single language ecosystem. The inconsistent inclusion of hidden directories 
across Python's own globbing libraries forces developers to memorize implementation-specific quirks and add defensive filtering logic, undermining the principle of predictable 
pattern matching.

\subsection{Performance Concerns and Compatibility Challenges}

Performance issues ($81$ posts) reveal significant scaling limitations in current implementations primarially focused on -- slow matching which accounts for $19$ posts ($23.5\%$), 
particularly affecting recursive operations ($6$ posts at $7.4\%$), highlighting inefficient traversal algorithms that lack pruning mechanisms or early termination capabilities.
The absence of optimization features like directory exclusion, and streaming results forces developers to implement workarounds that often worsen performance through 
unnecessary traversal and post-processing.








%% file: discussion.tex

Our empirical study revealed that globbing's critical deficiencies -- semantic fragmentation across languages, security vulnerabilities, and expressiveness limitations -- are rooted 
in a historical architectural choice. Specifically, the coupling of pattern semantics with system context. We propose a formal specification that codifies the common features 
in existing globbing system, providing a consistent and clear definition of behaviors, and provides a clean distinction between logical pattern meaning and system behavior 
for execution. 

\subsection{Separating Pattern Semantics from Execution}

The syntax of the proposed \textsc{GlobSpec} specification is defined using the context-free grammar (CFG) shown below. This grammar captures the core constructs 
common across existing globbing implementations while introducing new features to address key expressiveness gaps. 

\begin{lstlisting}[language=ebnf, caption={GlobSpec Grammar Specification}]
GLOB  ::= \PREFIX PATH SUFFIX\[uhlnis]*

PREFIX ::= URI_SCHEME
         | "/"
         | (@$\epsilon$@)

SUFFIX ::= "/"
         | (@$\epsilon$@)

PATH  ::= SEQ "/" PATH 
          | "**" "/" PATH 
          | SEQ 
          | "**"

OPTS  ::= SEQ "|" OPTS 
          | SEQ

SEQ   ::= TERM+

TERM  ::= literal
          | brange // e.g., [a-z]{3, 5}
          | *
          | "${" identifier "}"
          | "(" OPTS ")"
\end{lstlisting}

This grammar serves as the foundation for a two-layered architecture that distinctly separates \textbf{universal pattern semantics} from \textbf{execution behaviors}. The root 
production for a glob pattern is \texttt{GLOB}, which may be annotated with optional flags (\texttt{u}, \texttt{h}, \texttt{l}, \texttt{n}, \texttt{i}, \texttt{s}) to control execution behaviors without 
altering the core structure of the path itself. The physical path structure is defined by the \texttt{PATH} non-terminal, which recursively composes components (\texttt{SEQ})
using the forward slash (\texttt{/}) as a \emph{standardized} path separator where the, physical, character used for matching is controlled by the \emph{separator} (\texttt{s}) flag. 

The path can be made of up to two components: recursive wildcards (\texttt{**}) and concrete term sequences (\texttt{SEQ}). 
Concrete \texttt{SEQ} components can be simple terms made of literal characters, bounded range matches, wildcards, variable substitutions (\texttt{\$\{identifier\}}), or union constructs \texttt{(opt1|opt2)}.\\ 

\noindent
Sample patterns defined by this grammar include:
\begin{itemize}
\item \textbackslash\texttt{*.txt}\textbackslash is a single component path of a character sequence followed by the literal sequence of characters ``\texttt{.txt}''
\item \textbackslash\texttt{**/src/*.java}\textbackslash specifies a recursive prefix that ends in ``\texttt{src}'', then an immediate match of child files ending with ``\texttt{.java}"
\item \textbackslash\texttt{(cat|dog)*}\textbackslash defines the union of files beginning with ``\texttt{cat}'' or ``\texttt{dog}'', followed by any character sequence
\item \textbackslash\texttt{/docs/(*.log|*.tmp)}\textbackslash expresses the files matching either \texttt{*.log} or \texttt{*.tmp} patterns
\item \textbackslash\texttt{/\$\{user\}/docs/ff\_[0-9]\{3,\}.tmp}\textbackslash is a parameterized path \texttt{user} that matches files starting with ``\texttt{ff\_}'' followed by at 
least three digits and ending with ``\texttt{.tmp}''
\end{itemize}

These structural definitions are universal across all implementations. The grammar ensures that a pattern's logical intent is preserved regardless of the underlying platform, 
programming language, or filesystem characteristics, establishing a foundation for predictable and portable file matching operations.

The option flags control various environmental factors, as identified in this work, that are needed to adopt the logical structure of a path to the specifics of the environment 
and platform that it is executed on. In the \textsc{GlobSpec} architecture these flags can be set as part of the pattern itself, as global environmental settings, or as parameters to 
globbing function calls. The following flags are proposed to address the key sources of semantic fragmentation identified in our study:
\begin{itemize}
\item \textbf{(U)nicode}: Marks the glob as operating on \texttt{unicode} characters or by default the safe charset defined by the RFC3986 safe characters standard~\cite{urirfc}.
\item \textbf{(H)idden Files}: A \texttt{include\_hidden} flag determines whether hidden files (e.g., files starting with a dot on Unix systems) are considered during matching.
\item \textbf{(L)inks}: A \texttt{follow\_links} flag specifies whether symbolic links should be traversed during pattern matching.
\item \textbf{(N)egation}: A \texttt{negation} flag negates the resulting glob behavior and returns only non-matching paths.
\item \textbf{(I)nsensitive Matching}: A \texttt{ignore\_case} flag allows patterns to match files regardless of case (and unicode normalization).
\item \textbf{(S)eparator}: A \texttt{path\_separator} flag allows the user switch the character used as the path separator from forward slash \texttt{/} vs. backslash \texttt{\textbackslash{}}.
\end{itemize}

This architectural separation systematically addresses both the fragmentation and expressiveness limitations. The grammar eliminates semantic fragmentation by fixing core pattern structure 
and interpretation while the flags control execution details, like symlink following during recursion, and eliminates major classes of issues -- \eg this resolves cases where path separator 
ambiguity (18.7\% of cross platform failures) is eliminated through invariant parsing rules.

The expressiveness limitations documented in RQ4 find comprehensive solutions through \textsc{GlobSpec}'s composition mechanisms. The union, range complement, negation, and substitution 
rules enable advanced pattern operations previously requiring manual workarounds. This formalism resolves many persistent feature requests and workarounds seen in the developer 
posts and enables principled glob construction for cases like \texttt{(*.log|*.tmp)}. 
The substitution mechanism \texttt{\$\{}pattern\_name\texttt{\}} addresses 19.7\% of feature gap requests by enabling pattern reuse and composition. Negation support in the flags 
eliminates manual filtering workarounds that contribute to 23.5\% of performance issues, where developers currently process all files only to discard most during post filtering.
Crucially, \textsc{GlobSpec}'s regular language foundation enables linear time pattern checking through finite automata compilation.

\textsc{GlobSpec} transforms globbing from a world of \textbf{implicit, hidden behaviors} where identical pattern strings produce different results across contexts to a 
world of \textbf{explicit, controlled execution} where patterns maintain universal meaning through transparent, configurable interfaces.

This architectural shift enables reference implementations, pattern validation tools, cross platform testing suites, and security auditing capabilities previously infeasible 
with implementation defined globbing. By providing this formal foundation, \textsc{GlobSpec} addresses the reliability, security, and portability concerns quantified throughout 
our empirical study while establishing globbing as a verifiable software engineering primitive.

%% file: threats.tex

{\bf Glob Sampling Bias:} Our methodology for procuring the glob corpus faces two threats. First, our corpus is composed of globs sampled from publicly available projects on 
github in a fixed range of programming languages. If these sets are not sufficiently representative of the true distribution then our results may be biased. To counteract this 
risk, we selected a range of popular programming languages. Second, our selection methodology involved using a set of extraction regular expressions that may miss some 
uses of globs. To reduce this risk we randomly spot-checked and refined the extraction pipeline to ensure high coverage.

{\bf Discussion Bias and Categorization:} Our methodology for gathering and analyzing developer discourse on globs also contains two possible threats to validity. First, is 
possible sampling bias in the forums selected and the self-selection of developers that post on these sites. Our choice of the most popular forums should ensure that the results 
are representative of programming topics that are generally relevant. As data collection was keyword-based we manually evaluated both potentials for false positives and 
false negatives in our results. The second threat to validity of the results based on this analysis are the miss-classification of posts on topics. Our preliminary classification 
was based on keyword matches but was then hand refined to ensure that labeling was accurate.

%% file: related.tex

Studies such as~\cite{regexes-are-hard,type-system-regex,regex-usage-and-context,regex-comprehension,regex-test-in-wild,regexBugs,regex-evolution} explore the challenges developers 
face with pattern-matching tools, particularly regular expressions (regex). These studies highlight common pitfalls, such as incorrect pattern matching, performance issues, and 
security vulnerabilities like ReDoS (Regular Expression Denial of Service)~\cite{impact-of-ReDoS, rethink-regex}. While regex has been extensively studied, globbing has received 
less attention despite its widespread use in shell, scripting, build tools, and configuration management.

Our research extends these findings by focusing specifically on globbing's usability, portability, and shortcomings across programming environments. By analyzing real-world usage 
patterns and developer discussions, we identify gaps in globbing implementations and propose solutions to improve cross-platform compatibility and developer productivity. The 
semantics of pattern-matching tools vary significantly across programming languages and platforms. For example~\cite{disambiguation-of-regex} proposed an analysis for checking if 
a regex was sensitive to greedy/lazy match semantics. Globbing, as a specialized form of pattern matching, exhibits similar variability. These inconsistencies pose challenges 
for developers writing scripts. Our work builds on these studies by documenting the differences in globbing and proposing a standardized approach to improve consistency.

File system operations, including path manipulation and pattern matching, are fundamental to software development. Research such as~\cite{file-system-evolution} discusses the 
evolution of file systems and the semantics bugs across file systems. \cite{ACID} explored extending ACID semantics to file systems, emphasizing the importance of consistency 
and reliability in file handling. Globbing plays a critical role in file system operations, enabling developers to match and manipulate file paths. However, the lack of standardization 
in globbing implementations leads to portability issues, particularly when scripts are moved between Unix-like systems and Windows. Our research addresses these challenges by 
analyzing globbing behavior across platforms and proposing enhancements to improve reliability and portability. 

Security vulnerabilities in pattern-matching tools, such as path traversal attacks and ReDoS~\cite{ReDoScase} in regular expressions, have been widely documented. Globbing is not 
immune to these issues. Attacks in the wild that use improper validation of glob patterns, can lead to path traversal attacks, or globbing introduced performance 
attacks~\cite{insecureparsing, heapoutofbound,stackexhaustion}. Our research examines these security implications and proposes best practices for secure glob patterns. 

Cross-platform compatibility is a recurring challenge in software development. These challenges are exacerbated by inconsistent behavior of globbing, such as differences in 
case sensitivity and path separator handling, \eg~\cite{caseinsensitive} and \cite{pathescape}. Our study builds on this work by empirically analyzing globbing behavior across 
multiple platforms and programming languages, identifying common sources of incompatibility, and proposing solutions to improve portability.